\documentclass[sn-mathphys-num]{sn-jnl}

\usepackage{graphicx}%
\usepackage{multirow}%
\usepackage{amsmath,amssymb,amsfonts}%
\usepackage{amsthm}%
\usepackage{mathrsfs}%
\usepackage[title]{appendix}%
\usepackage{xcolor}%
\usepackage{textcomp}%
\usepackage{manyfoot}%
\usepackage{booktabs}%
\usepackage{algorithm}%
\usepackage{algorithmicx}%
\usepackage{algpseudocode}%
\usepackage{listings}%

\theoremstyle{thmstyleone}%
\newtheorem{theorem}{Theorem}
\newtheorem{proposition}[theorem]{Proposition}%

\theoremstyle{thmstyletwo}%

\theoremstyle{thmstylethree}%

\def\tr{\mathop{\rm tr}\nolimits}
  \newtheorem{lemma}{Lemma}

\usepackage{array}

\begin{document}

\title{Vector field dynamics: field equations and energy tensor} 


\author*[1]{\fnm{Roberto} \sur{Dale}}\email{rdale@umh.es}

\author[2]{\fnm{Alicia} \sur{Herrero}}\email{aherrero@mat.upv.es}
\equalcont{These authors contributed equally to this work.}

\author[3,4]{\fnm{Juan Antonio} \sur{Morales-Lladosa}}\email{antonio.morales@uv.es}
\equalcont{These authors contributed equally to this work.}

\affil[1]{\orgdiv{Departamento de Estad\'istica, Matem\'atica e Inform\'atica and Centre of Operations Research (CIO)}, \orgname{Universidad Miguel Hern\'andez},\orgaddress{
 \city{ Elx}, \postcode{03202},
\country{ Spain}}}

\affil[2]{\orgdiv{Instituto de Matem\'atica Multidisciplinar and Departamento de Matem\'atica Aplicada}, \orgname{Universitat Polit\`ecnica de Val\`encia}, \orgaddress{
\city{Valencia}, \postcode{46022}, 
\country{Spain}}}

\affil[3]{\orgdiv{Departament d'Astronomia i Astrof\'{\i}sica}, \orgname{Universitat de Val\`encia}, \orgaddress{
\city{Burjassot}, \postcode{46100}, 
\country{Spain}}}

\affil[4]{\orgdiv{Observatori Astron\`omic}, \orgname{Universitat de Val\`encia}, \orgaddress{
\city{Paterna}, \postcode{46980}, 
\country{Spain}}}


\abstract{Relativistic field theory for a vector field on a curved space-time is considered assuming that the Lagrangian field density is quadratic and contains field derivatives of first order at most. 
By applying standard variational calculus, the general Euler-Lagrange equations for the field are derived and the existence of a conserved current is achieved. The field equations are also analysed from an eikonal-like point of view. The Hilbert energy-momentum tensor of the field is also derived and the influence of each one of the irreducible pieces appearing in the Lagrangian is studied. Particular values of the free parameters allow to retrieve known results.
}




\maketitle


\section{Introduction}
\label{sec:1}

Space-time vector fields are essential tools in Physics. In the standard model of Particle Physics, the fundamental interaction between elementary fermions (quarks and leptons) are mediated by twelve vector fields (eight gluons and four electroweak spin 1 bosons) \cite{Peacock,Nagashima}. Lie algebras generators are used to implement diverse physical symmetries, both in Relativistic Quantum Mechanics and in General Relativity \cite{Szekeres}. 
Moreover, beyond the standard model of particle physics, a self-interacting vector field may describe the $U(1)$ symmetry breaking and its implications in Cosmology when the field is coupled to gravity (see, for instance, Refs. \cite{Tasi2014, hei14}). A vector field is the architect for a tensor-vector theory of gravitation like, for example, the one introduced in \cite{wil93,wil06} or its extensions (see, for instance \cite{beltran13,jim16} and \cite{Lavinia2019} for a review).
The search for dynamical and gravitational predicted imprints that could be attributed to, otherwise unobserved,  vector fields is a current issue in Astrophysics and observational Cosmology \cite{dal15,dal17,dal18,dal19}.  

This work is focused on the dynamics of a ``test vector field'' $\xi$. This task involves (i) the analysis of the field equations for $\xi$ and (ii) the description of its energy-momentum tensor. The terminology ``test field'' underlines the fact that no specific metric theory of gravitation is considered along this work. Then, any test field characteristic property works in any curved space-time background and constitutes the starting point for studying self-gravitating vector fields. This last scenario involves, in addition, the Einstein field equations sourced by the energy content defined by $\xi$, and the coupling of the dynamical evolution of the field itself. 

In this paper, the obtained results will be presented as general as possible in order to open the possibility of applications in extended areas of Physics, including General Relativity. For instance, the field energy tensor could be studied from an algebraic point of view and characterised according to the Churchill-Pleba\'nski classification \cite{Churchill,Plebanski-64} of a space-time symmetric 2-tensor and the restriction derived from the usual energy conditions introduced by Pleba\'nski (see Refs. \cite{Plebanski-64,BCM-92}). The $3+1$ (space plus time) decomposition of the field equation and the energy tensor of $\xi$, with respect to a space-time observer $u$  (unit, future oriented time-like vector field), and their physical interpretation are other possible applications of our work. In particular, when $\xi$ is time-like and $u$ is the observer associated to $\xi$, the decomposition of its covariant derivative, $\nabla u$, relatively to $u$ provides the kinematic properties of $\xi$ (acceleration, shear, expansion and vorticity), which are involved in the field equations of $\xi$ and help to interpret its energy tensor. Moreover, the general approach to self-gravitating vector fields involving the $3+1$ decomposition of the coupled system of the Einstein equations and the vector field equations, the former sourced by the energetic content of the field $\xi$ itself, could also be investigated. 

The terminology time-like, light-like (null) or space-like for a vector $\xi$ will be used to express the fact that $\xi^2\equiv g(\xi, \xi)$ is negative, zero or positive, respectively, when the signature convention $(-,+,+,+)$ is taken for the space-time metric $g$. However, no particular election for the metric signature has been used along this work. 

The paper is structured according to this plan. Notation, main definitions and sign conventions are established in Sec. \ref{sec:2}. The Lagrangian density $L$ for a vector field theory and the corresponding Euler-Lagrange equations are introduced in Sec. \ref{sec:3}, pointing out  the fundamental terms in $L$.
Next, Sec. \ref{sec:5} deals with the eikonal-like decomposition of the dynamics of a vector field.
Following the Hilbert variational prescription, in Sec. \ref{sec:4}, the energetic content carried out by a vector field is obtained piece to piece, according to the Lagrangian decomposition in invariant summands. Finally, the results of this paper are discussed and summarised in Sec. \ref{sec:6}.
To gain conciseness, the main results are presented as statements under the form of brief propositions and lemmas.


\section{Notation and conventions}
\label{sec:2}

First of all, some remarks about the notation and some useful identities used further in this paper are given:

(i) The space-time metric determinant, ${\rm g} \equiv \det g < 0$, fulfills the relations:
 \begin{equation}\label{detg}
 {\rm g} = g_{\mu\nu}m^{\mu\nu} \qquad {\rm and
 } \qquad \frac{\partial {\rm g}}{\partial g_{\mu\nu}} 
 = m^{\mu\nu} = {\rm g} \, g^{\mu\nu} \, ,
 \end{equation}
where $m^{\mu\nu}$ is the minor (with its sign) of the matrix metric element $g_{\mu\nu}$, in a given basis, and $g^{\mu\nu}$ is the inverse of the metric, used to raise and lower indexes, respectively.  Other related expressions, which will be used in Section 6 to deduce Eq. (\ref{T-Hilbert-bis}), are:
 \begin{equation}\label{lng}
 g^{\mu\nu} = \frac{\partial{\ln  {(\rm - g)}}}{\partial g_{\mu\nu}}\, \qquad {\rm and
 } \qquad
 \Gamma^\alpha_{\alpha\rho} = \partial_\rho \ln \sqrt{\rm - g} \, ,
\end{equation}
 with $\Gamma^\alpha_{\alpha\rho} = g^{\alpha \mu} \Gamma_{\alpha\rho . \mu}$ (summation  over contracted repeated indexes is understood)  and  
 $$
 \Gamma_{\alpha\rho . \mu} = 
 \frac{1}{2}(\partial_\alpha g_{\rho\mu} + \partial_\rho g_{\alpha\mu} -  \partial_\mu g_{\alpha\rho})
 $$ 
 as the Christoffel symbols, defined from the first derivatives of the metric field components, $g_{\mu\nu} (x^\alpha)$, in a coordinate system $\{x^\alpha\}$, which are denoted by 
 $$ 
 g_{\mu\nu,\alpha} \equiv \partial_\alpha g_{\mu\nu} 
 = \frac{\partial g_{\mu\nu}}{\partial x^\alpha}.
 $$

 (ii) Let $\xi$ be a smooth vector field, defined on a space-time domain $D$ with metric $g$ and Levi-Civita connection $\nabla$. 
 From the covariant derivative of $\xi$, $\nabla \xi$, and its transpose, $^{\rm t} \nabla \xi \equiv (\nabla \xi)^{\rm t}$, we can always build two 2-tensors, $S$ and $F$ defined by:
 \begin{eqnarray}
S &\equiv & \nabla \xi + \vspace{0.5mm}  ^{\rm t}\nabla \xi = {\cal L}_\xi g \, , \label{def-S}\\
F  & \equiv & \nabla \xi -  \vspace{0.5mm} ^{\rm t} \nabla \xi = d \xi\, \label{def-F}.
\end{eqnarray}
where ${\cal L}_\xi$ stands for the Lie derivative along $\xi$ and $d$ for the exterior derivative, here acting on the 1-form metrically equivalent to $\xi$, which is also denoted by $\xi$, that is $g(\xi, \cdot ) \equiv\xi$.
By using index notation the 2-tensors $S$ and $F$ are:
\begin{eqnarray}
S_{\mu\nu} & = & \nabla_\mu \xi_\nu + \nabla_\nu \xi_\mu = ({\cal L}_\xi g)_{\mu\nu} \, , \label{def-S2}\\
F_{\mu\nu}  & = & \nabla_\mu \xi_\nu - \nabla_\nu \xi_\mu = (d \xi)_{\mu\nu} \label{def-F2}
\end{eqnarray}
respectively, where $\nabla_\mu \xi_\nu\equiv (\nabla\xi)_{\mu\nu}=\partial_\mu\xi_\nu - \Gamma_{\mu\nu}^\rho \xi_\rho$.

(iii)  Given $x$ and $y$, both being vector or covector fields, the symbol $\tilde\otimes$ stands for their symmetrised tensorial product, that is, $x \tilde \otimes y \equiv x \otimes y + y \otimes x$. 

(iv) Given $P$ and $Q$ second order tensors, the tensor $P \times Q$ denotes its matrix product, or contraction of adjacent indices, that is 
$$(P \times Q)_{\mu}\, ^{\nu} = P_{\mu \rho} \, Q^{\rho\nu}.$$ 
In particular, $P^2=P\times P$.
The trace of $P$ is $\tr P=g^{\mu\nu}P_{\mu\nu} = P^\mu_\mu$. Thus, for the 2-tensors $S$ and $F$ in (\ref{def-S}) and (\ref{def-F}), 
\begin{equation} \label{trS}
\tr S = 2\, \nabla_\mu \xi^\mu \equiv 2 \, \nabla \cdot \xi,
\end{equation}
where 
\begin{equation} \label{divergencia}
\nabla \cdot \xi\equiv\nabla_\mu \xi^\mu=\displaystyle{\frac{1}{\sqrt{ -{\rm g} }}\partial_\mu(\sqrt{-{\rm g}}\ \xi^\mu)}
\end{equation} 
is the covariant divergence of $\xi$. Moreover,
\begin{eqnarray}
\tr S^2  & = & \tr (S \times S)  = S_{\mu\nu} S^{\mu\nu} \, ,  \label{trS2}  \\     
\tr F^2  & = & \tr (F \times F)  = - F_{\mu\nu} F^{\mu\nu} \, , \label{trF2}
\end{eqnarray}
because $S$ is symmetric and $F$ is antisymmetric, and then $\tr(S \times F) = \tr (F \times S) = 0$. %
 
(v) The divergence of a 2-tensor $T$ is denoted by $\nabla \cdot T$ (in components,  $(\nabla \cdot T)^\nu = \nabla_\mu T^{\mu\nu}$). The interior or contracted product by $\xi$ is denoted by $i(\xi)$. 
For instance, if $T$ is a covariant $2$-tensor, then $(i(\xi) \nabla T)_{\mu\nu} = \xi^\alpha \nabla_\alpha T_{\mu\nu}$ is the covariant derivative of $T$ along $\xi$, that is $i(\xi)\nabla  T = \nabla_\xi T$.  On the other hand, $[i(\xi)T]_\nu = \xi^{\mu} T_{\mu\nu}$ is a covector but if $T$ is a mixed $2$-tensor,  $[i(\xi)T]^\nu = \xi^{\mu} T_{\mu}^{\;\,\nu}$ is a vector. The double contraction of $T$ with $\xi$ is denoted by $i^2(\xi) T \equiv i(\xi) i(\xi) T = T(\xi, \xi) = T_{\mu\nu} \xi^\mu \xi^\nu$.

(vi) By convention, the sign of the Riemann curvature tensor is taken according with the Ricci identities, which are written as 
\begin{equation}\label{RicId}
\nabla_\mu \nabla_\nu \xi_\beta - \nabla_\nu \nabla_\mu \xi_\beta = \xi^\alpha R_{\alpha \beta \nu \mu} \, .  
\end{equation}
The Ricci tensor of $g$, $Ric \equiv Ric(g)$,  is defined as the contraction of the first and third index of the curvature (in components $R_{\alpha\beta} = R^{\mu}_{\ \alpha \mu \beta}$), and the scalar curvature is $R \equiv \tr Ric$. Moreover, the Ricci identities for 2-tensors have the expression:
\begin{equation}\label{RicId2}
\nabla_\mu \nabla_\nu T_{\alpha\beta} - \nabla_\nu \nabla_\mu T_{\alpha\beta} = R^\rho_{\ \alpha \nu \mu} T_{\rho\beta} +  R^\rho_{\ \beta \nu \mu} T_{\alpha\rho} \, .  
\end{equation}

%

\section{Vector field dynamics}   
\label{sec:3}

This section is mainly devoted to obtaining the differential equation that a vector field obeys, following the usual approach of the relativistic field theory in a curved space-time. Moreover, a detailed comparison with previous studies, carried out by other authors, is also provided. 

In order to avoid nonlinearity in first and second derivatives of the field equation, we shall consider a Lagrangian density of the  vector field  containing at most first-order derivatives (of the field). This density is constructed from quadratic invariants of the field and its derivatives. This assumption allows us to derive a wave-type equation, that is, a hyperbolic main part of the equation. Furthermore, as we discuss in the concluding section, higher order invariants generically lead to a field equation with nonlinear terms formed by products of the first and second-order derivatives.


\subsection{Euler-Lagrange equation for a vector field $\xi$}
\label{sec:3a}

For a vector field $\xi$ in a curved geometry, we are faced with the dynamic field equation that $\xi$ obeys. Variational calculus from a density Lagrangian leads to the corresponding Euler-Lagrange equations for $\xi$, whose relevant differential terms depend on the hypothesis that takes place in the construction of $L$. 

Let us consider the functional:
\begin{equation}\label{funcional}
I = \int_D L \,  \sqrt{- {\rm g}} \, \, d^4x, \quad \quad {\rm g} \equiv \det g
\end{equation}
where $L$ is the quadratic Lagrangian density on $D$, defined from the vector field $\xi$ and its first derivatives as
\begin{equation}\label{lagra}
L = a \tr S^2 + b \tr F^2 + c  \, (\tr S)^2 + (\mu + \nu R) \, \xi^2 \, ,
\end{equation}
with $a, b, c, \mu, \nu \in \mathbb{R}$ and $R$ the scalar curvature of $g$. 

Note that our Lagrangian density is defined using the scalars $\tr S^2$, $ \tr F^2$ and $\tr S$ without consider terms on the Ricci tensor of the form $i^2(\xi) Ric$. Later on, it will be seen that this Ricci terms can be rewritten in terms of the scalars used in our definition of the Lagrangian density.

By requiring $L$ to be stationary under  variations of $\xi$ and its first derivatives (vanishing on $\partial D$, the  boundary of $D$), the Euler-Lagrange equations \cite{Hawking-Ellis} are written as:
\begin{equation}\label{Euler-Lagrange}
\nabla_\alpha \frac{\partial L}{\partial \nabla_\alpha \xi_\beta} - \frac{\partial L}{\partial \xi_\beta} = 0.  
\end{equation}

Notice that considering an extra term proportional to $\tr S$ in expression (\ref{lagra}) of $L$ is irrelevant, since
\begin{equation}\label{Ldiv}
\frac{\partial \tr S}{\partial \nabla_\alpha \xi_\beta} = 2 g^{\alpha\beta} \, , 
\end{equation}
and $\nabla g = 0$. Then, such a term does not contribute to the field equation (\ref{Euler-Lagrange}), in accordance with the application of the Gauss theorem for the case $\tr S=2\nabla\cdot \xi$ and the  assumed boundary conditions. As a matter of consistence, the energy tensor associated with this extra term has to be zero (see Proposition {\ref{L1} in Subsection \ref{sec:4b}). 

Now, from the definition of $S$ and $F$ in (\ref{def-S2}) and (\ref{def-F2}), respectively, the following relations for the derivatives of the different summands in expression (\ref{lagra}) of $L$ are deduced:
\begin{eqnarray} \label{sumand1}
\frac{\partial \tr S^2}{\partial \nabla_\alpha \xi_\beta} & = & 4 \, S^{\alpha\beta}, \\  \label{sumand2}
\frac{\partial \tr F^2}{\partial \nabla_\alpha \xi_\beta} & = &  - 4 \, F^{\alpha\beta}, \\ \label{sumand3}
\frac{\partial (\tr S)^2}{\partial \nabla_\alpha \xi_\beta} & = & 4 \, \frac{\partial (\nabla \cdot \xi)^2}{\partial \nabla_\alpha \xi_\beta} = 8 \, (\nabla \cdot \xi ) \, g^{\alpha\beta}, \\ 
\label{sumand4}
\frac{\partial \xi^2}{\partial \xi_\beta} & = & 2 \, \xi^{\beta}.
\end{eqnarray}
Substituting these expressions in Eq. (\ref{Euler-Lagrange}), it becomes:
\begin{equation}\label{Eq-camp-1}
a \, \nabla \cdot S - b \, \nabla \cdot F + 2\, c \, d \, (\nabla \cdot \xi) - \frac{1}{2} (\mu + \nu R) \xi = 0 \, ,
\end{equation}
which is the field equation that the vector field $\xi$ satisfies. 

Taking into account  the Ricci identities (\ref{RicId}) for $\xi$, the divergence summands, $\nabla \cdot S$ and  $\nabla \cdot F$, might be conveniently expressed in terms of the field and its derivatives as follows: 
\begin{eqnarray}\label{divS}
 \nabla \cdot S & = & \Delta \xi  + d (\nabla \cdot \xi) + i(\xi) Ric \, , \\ \label{divF}
 \nabla \cdot F & = & \Delta \xi  -  d (\nabla \cdot \xi) - i(\xi) Ric \, ,
\end{eqnarray}
where $\Delta$ stands for the Laplace-Beltrami operator,  $(\Delta \xi)_\alpha \equiv g^{\mu \nu} \nabla_\mu \nabla_\nu \xi_\alpha$. Then, redefining the coefficients in Eq. (\ref{Eq-camp-1}) as 
\begin{equation}\label{parameters}
\alpha \equiv a-b, \quad \beta \equiv a + b, \quad \gamma \equiv a + b + 2c \quad {\rm and} \quad \rho \equiv - \frac{1}{2} (\mu + \nu R),
\end{equation}
we have the following result.

\begin{proposition}
For a Lagrangian of the form (\ref{lagra}) associated to the field $\xi$, the Euler-Lagrange equation is
\begin{equation} \label{Eq-camp-2}
\alpha \,  \Delta \xi + \beta \,  i(\xi)Ric + \gamma \, d \,  (\nabla \cdot \xi) + \rho \, \xi = 0,
\end{equation}
where $\alpha$, $\beta$, $\gamma$ and $\rho$ are given in (\ref{parameters}).
\end{proposition} 
The term $\rho \, \xi$ acts as a generalized (coupled to gravity via the scalar curvature $R$) mass term in the field equations, which comes from the last summand in equation (\ref{lagra}).

Related to the hyperbolicity (strong and weak) analysis and stability of this field equation (\ref{Eq-camp-2}) for $\mu=\nu=0$,
in the space ${\mathbb R}^3$ of real parameters $\alpha, \beta, \gamma$, we can closely follow the procedure presented in Ref. \cite{FengGaWi-19}. In the generic case $\alpha\neq 0$, we get that Eq. (\ref{Eq-camp-2}) is strongly hyperbolic only for $\gamma=0$ and weakly hyperbolic if $\gamma \neq -\alpha$.

Then, Eq. (\ref{Eq-camp-2}) is the field equation for the vector field $\xi$ in any metric $g$. This equation will be called here the {\em Proca-Taubes-Will equation} since it 
reduces to the field equations considered by these three authors when particular values of the parameters are chosen, as it will be seen in the next subsection.

\subsection{Discussion about the field equation.}
\label{sec:3b}

Firstly, the previous field equation (\ref{Eq-camp-2}) will be compared with the Proca field equation \cite{Proca-1936}. To begin with, let us consider the Proca Lagrangian density, 
$$
L_{\cal P} = \frac{1}{4}\tr F^2 - \frac{\mu_p^2}{2} \, \xi^2,
$$
being $\mu_p$ the Proca mass. This $L_{\cal P}$ clearly follows from (\ref{lagra}) taking $a = c= \nu =0, b=\displaystyle{\frac{1}{4}}$ and $\displaystyle \mu=-\frac{\mu_p^2}{2}$. From these values of the parameters, Eq. (\ref{Eq-camp-1}) becomes 
\begin{equation}\label{Proca-1}
\nabla \cdot F - \mu_p^2 \, \xi = 0, 
\end{equation}
which is the Proca field equation. Since $F$ is antisymmetric, this equation implies that the Proca field is divergence free, $\nabla \cdot \xi = 0$, which is not a gauge condition. The Proca field itself defines a conserved current when $\xi$ is time-like. Then Eq. (\ref{Eq-camp-2}) reduces to
\begin{equation}\label{Proca-2}
\Delta \xi  - i(\xi)Ric - \mu_p^2 \, \xi = 0 \, .
\end{equation}
%
Regarding (\ref{Eq-camp-2}), notice that the parameter $\gamma=b=\displaystyle{\frac{1}{4}}$ does not appear since the field is divergence free.

Let us continue with the Almost-Killing Equation (AKE): 
\begin{equation}\label{AKE}
\Delta \xi + i(\xi)Ric + (1-\lambda) d \,  (\nabla \cdot \xi) = 0 \, ,
\end{equation}
introduced by Taubes \cite{Taubes-78} to extend the concept of exact space-time isometry. Clearly, this expression is recovered from Eq. (\ref{Eq-camp-2}) taking
$\alpha=1$, $\beta=1$, $\gamma=1-\lambda$ and $\rho=0$ (that is, 
$a = 1$, $b = 0$, $2c = - \lambda$ and $\mu = \nu =0$). Related to this equation, in Ref. \cite{BoCaPa-05}, the authors vindicate the AKE as a useful gauge condition to be implemented in the 3+1 formulation of the Einstein field equations. Moreover, Ref. \cite{FengGaWi-19} contains a general analysis of the AKE; in particular, the AKE equation is strongly hyperbolic iff $\lambda = 1$,  that corresponds to the differential equation, $\Delta \xi + i(\xi)Ric = 0$, for a generator $\xi$ of an harmonic transformation, under which the Levi-Civita connection satisfies $g^{\mu\nu}{\cal L}_\xi \Gamma^\alpha_{\mu\nu} = 0$  (see Ref. \cite{Nouhaud-72}). The definition of the Lie derivative for a linear connection (${\cal L}_\xi\Gamma^\alpha_{\mu\nu}$) can be found in \cite{Yano}. 

Next, the Will equation coming from the Will Lagrangian density,  
 \begin{equation}\label{Will-lagra}
 L_{\cal W} = \epsilon \, \tr F^2 + \eta \,  i^2(\xi) \, Ric +  \tau \tr (\nabla \xi \times \vspace{0.5mm}^{\rm t}\nabla \xi) + \omega \, R \, \xi^2 \, ,
 \end{equation}
with $\epsilon, \eta, \tau$ and $\omega$, real parameters, is analysed. Originally, $L_{\cal W}$ was introduced to build vector-tensor theories of gravitation, by extending the Hilbert variational formulation of the Einstein' General Relativity (see Refs. \cite{wil93,wil06}).  In this context, the full Lagrangian density is written as $L = R + L_{\cal W} + L_m$, where $L_m$ stands for the matter term. From  the Lagrangian (\ref{Will-lagra}),  the field $\xi$ obeys the Will equation:
\begin{equation}\label{Will-1}
\epsilon \, \nabla \cdot F + \frac{\eta}{2} \, i(\xi) \, Ric - \frac{\tau}{2} \, \Delta  \xi +  \frac{\omega}{2} \, R \, \xi = 0.
\end{equation}

Taking into account the general expression (\ref{divF}) for $\nabla \cdot F$, this equation (\ref{Will-1}) is written as
 \begin{equation}\label{Will-2}
\left(- \epsilon + \frac{\tau}{2}\right) \, \Delta \xi  + \left(\epsilon - \frac{\eta}{2}\right) \, i(\xi)Ric
+ \epsilon \, d \, (\nabla \cdot \xi) - \frac{\omega}{2} \, R \,  \xi = 0, \, 
\end{equation}
which can be obtained from Eq. (\ref{Eq-camp-2}) making
 \begin{equation}\label{corres1}
\alpha = - \epsilon + \frac{\tau}{2},\; \beta=\epsilon - \frac{\eta}{2}, \; \gamma=
 \epsilon, \; \rho=-\frac{\omega}{2}R
\end{equation}
that is, taking form Eq. (\ref{lagra})
 \begin{equation}\label{corres2}
a = \frac{1}{4} (\tau - \eta), \; b = \epsilon - \frac{1}{4} (\tau + \eta), \;   c =\frac{\eta}{4}, \; \mu = 0, \; \nu = \omega.
\end{equation}
 
Notice that, up to the addition of a Proca-like term, $\mu \, \xi^2$,  the Lagrangian densities (\ref{lagra}) and (\ref{Will-lagra}) only differ by a full divergence term, and then, can be considered equivalent with respect to the obtention of the field equation from a variational principle with the assumed boundary conditions. To be more specific, 
\begin{eqnarray}\label{lagra-Will}
L_{\cal W} & = & \frac{1}{4} (\tau - \eta) \, \tr S^2 + [\epsilon -   \frac{1}{4} (\tau + \eta)] \, \tr F^2 + \eta \, (\nabla \cdot \xi)^2 \nonumber \\
& & +\, \omega \, R \, \xi^2 + \eta \, \nabla \cdot {\cal B} \, ,
\end{eqnarray}
where $ {\cal B} \equiv \nabla_{\xi} \xi - (\nabla \cdot \xi) \, \xi$. 
Now, since the following contracted Ricci identity
\begin{equation}\label{i2Ric}
i^2(\xi) Ric =  \nabla \cdot {\cal B} + (\nabla \cdot \xi)^2 - \tr (\nabla \xi \times \nabla \xi)
\end{equation}
is obtained by contraction of Eq. (\ref{RicId}) and, taking into account the expressions
\begin{eqnarray}
\tr (\nabla \xi \times \nabla \xi) & = & \frac{1}{4} (\tr S^2 + \tr F^2) \, ,  \label{trsi} \\  \label{trsi2} 
\tr (\nabla \xi \times \vspace{0.5mm}^{\rm t}\nabla \xi) & = & \frac{1}{4} (\tr S^2 - \tr F^2) ,
\end{eqnarray}
Eq. (\ref{lagra-Will}) is derived from Eq. (\ref{Will-lagra}).
The term $i^2(\xi) Ric$ in (\ref{Will-lagra}) is usually interpreted as a minimal coupling gravitational term. It is worthwhile to remark that the explicit introduction of this term in (\ref{lagra}) is a matter of convenience due to it could be incorporated by renaming the corresponding coefficients in (\ref{lagra}) using (\ref{trsi}) and (\ref{i2Ric}).

Finally, other cases can be obtained from expression (\ref{Eq-camp-2}). For $\alpha=1$, $\beta=-1$, ${\displaystyle\gamma=\frac{-2\epsilon}{2\epsilon-\eta}}\neq 0$ and $\rho=0$ (that is, $a=0$, 
$b=-1$, $2c={\displaystyle\frac{-\eta}{2\epsilon-\eta}}$ 
and $\mu=\nu=0$ if we look at Eq. (\ref{Eq-camp-1})), this equation reduces to the {\em Dale-S\'aez equation} \cite{dal14}: 
$$
\nabla\cdot F-\frac{\eta}{2\epsilon - \eta}\, d\, (\nabla\cdot\xi)=0
$$
or equivalently,
 \begin{equation}\label{DS-eq}
 \Delta \xi  - i(\xi)Ric - \frac{2 \epsilon}{2 \epsilon - \eta} \, d \, (\nabla \cdot \xi) = 0, \, 
\end{equation}
which can also be obtained from the Will equation (\ref{Will-2}) making $\tau = \eta \neq 2 \epsilon$ and $\omega = 0$.  %
Different aspects of a vector-tensor theory founded in the coupling of (\ref {DS-eq}) with the Einstein field equations has been analysed in Refs. \cite{dal14,dal09,dal12a,dal12b}. 

The case $\tau = 0,  \epsilon = -1$ in the Will Lagrangian density in (\ref{lagra-Will}) with the addition of  a Proca-like term $\mu \, \xi^2$, has also been treated in the literature. In such a case, we obtain 
\begin{equation}\label{Will-1-BoHa}
- \nabla \cdot F + \frac{\eta}{2} \, i(\xi) \, Ric +  \frac{1}{2}(\mu + \omega \, R) \, \xi = 0 \, ,  
\end{equation}
in consonance with  Eq. (\ref{Will-1}).  Notice that the precedent equation  (\ref{Will-1-BoHa})  and Eq. (4) in Ref. \cite{boh07} differ quite lightly, that is, the numeric coefficients of the term $\nabla \cdot F$ between both equations differ in a factor 2. However, this minor discrepancy should be taken into account in numerical or analytical expressions derived in \cite{boh07}. Furthermore, according with  Eq. (\ref{Will-2}), in this case,  we obtain that $\xi$ satisfies the equation:
\begin{equation}\label{Will-2-BoHa}
 \Delta \xi  - \left(1+ \frac{\eta}{2}\right) \, i(\xi)Ric - d \, (\nabla \cdot \xi) - \frac{1}{2} (\mu + \omega \,  R ) \xi = 0 \, 
\end{equation}
which corresponds to the case $\alpha=1$, $\displaystyle\beta=-1- \frac{\eta}{2}$, $\gamma=-1$ and $\displaystyle\rho=- \frac{1}{2}(\mu + \omega \,  R )$ (that is, $\displaystyle a=-\frac{\eta}{4}$, $\displaystyle b=-1 -\frac{\eta}{4}$, $c=\displaystyle\frac{\eta}{4}$, and $\mu = \mu$ and $\nu =\omega$).

In Table \ref{tab1} the previous discussion is summarised indicating the values of the parameters in the field equation (\ref{Eq-camp-2}) that recovers some of the known equations from the literature. 
\begin{table}[!h]
    \centering
      \caption{Summary of the values of the parameters in equation (\ref{Eq-camp-2}) to recover different field equations: Proca, Almost Killing Equation (AKE), Will and Dale-S\'aez (D-S). In the Proca case, the scalar function $\rho$ is related to the Proca mass $\mu_p$.} 
    \label{tab1}
     \begin{tabular}{ccccc} 
  \toprule
               & $\alpha$ & $\beta$ & $\gamma$ & $\rho$  \\ 
               \midrule
         Proca & $\displaystyle-\frac{1}{4}$ & $\displaystyle\frac{1}{4}$ & $\displaystyle\frac{1}{4}$ &  ${\displaystyle\mu=-\frac{\mu_p^2}{2},\, \nu=0 \rightarrow \rho=\frac{\mu_p^2}{4}}$ \\ \midrule 
         AKE & 1 & 1 & $1-\lambda$ & $\mu=\nu=0 \rightarrow \rho=0$  \\   \midrule
         Will & $-\epsilon +{\displaystyle\frac{\tau}{2}}$ & $\epsilon-{\displaystyle\frac{\eta}{2}}$ & $\epsilon$ & $\mu=0,\, \nu=\omega \rightarrow \rho=-{\displaystyle\frac{\omega}{2}R}$  \\  \midrule
         D-S & 1 & $-1$ & ${\displaystyle\frac{\-2\epsilon}{2\epsilon-\eta}}$ & $\mu=\nu=0 \rightarrow \rho=0$  \\ 
\botrule
    \end{tabular}
\end{table}

\subsection{Further considerations from the field equations}

The term $\nabla\cdot F$ in the field equation (\ref{Eq-camp-1}) allows to obtain, by derivation of this equation, a conserved current $J$. Effectively, due to the fact that $F$ is antisymmetric, the equation (\ref{RicId2}) leads to $\nabla\cdot(\nabla\cdot F)=0$ after the double contraction of this Ricci identity for $F$. Then, taking the divergence in the field equation (\ref{Eq-camp-1}) we have:
\begin{equation}\label{corrent}
\nabla\cdot J=0
\quad {\rm with} \quad J= 
b\, \nabla\cdot F.
\end{equation}
So, when $J$ is time-like, it could be interpreted as a conserved current.

The above expression for $J$ can be further simplified taking into account the  expression (\ref{divF}) for $\nabla\cdot F$:
\begin{equation}\label{corrent1}
 J=b\,\left( \Delta\xi-d(\nabla\cdot \xi)-i(\xi)Ric\right).
\end{equation}

Substituting the field equation (\ref{Eq-camp-2}), we get the following result:

\begin{proposition}
A conserved current $J$ derived from the field equation (\ref{Eq-camp-1}) when $\alpha\neq 0$ is given by:
\begin{equation}\label{corrent2}
\begin{array}{ccl}
 J & = & {\displaystyle\left[ 1-\frac{1}{2}\left(1+\frac{\beta}{\alpha}\right)\right]\rho\ \xi +\frac{1}{2}(\alpha+\beta)\left(1-\frac{\beta}{\alpha}\right)i(\xi)Ric }\\ [0.7cm]
&  &  {\displaystyle+ \left[\frac{\gamma}{2}\left(1-\frac{\beta}{\alpha}\right)+\frac{\alpha-\beta}{2}\right]d(\nabla\cdot \xi)}. 
\end{array}
\end{equation}
\end{proposition}
Notice that, when $\alpha\neq 0$, three contributions appear in (\ref{corrent2}) due to the terms $\xi$, $i(\xi)Ric$ and $d(\nabla\cdot \xi)$. Moreover,  if $\beta=\alpha\neq 0$, the resulting current identically vanishes since $b=0$.
And when $\beta=-\alpha\neq 0$,  a simplified situation follows,
\begin{equation}
J=\rho\ \xi +(\gamma-\beta)d(\nabla\cdot\xi),
\end{equation}
which has been analysed in Ref. \cite{wil93,wil06,dal12b} taking $\beta=-1$, $\gamma=-\displaystyle\frac{2\epsilon}{2\epsilon-\eta}$, $\rho=0$ and corresponds to 
the current,
\begin{equation}
J=-\frac{\eta}{2\epsilon-\eta}d(\nabla\cdot\xi).
\end{equation}

On the other hand, in an arbitrary space-time, the expression
\begin{equation}\label{moment1}
\Pi^\mu \equiv \frac{\partial L}{\partial \nabla_0 \xi_\mu}
\end{equation}
provides a logical generalization for the definition of conjugate momentum of the vector field $\xi$. In a flat space-time field theory, it is usually obtained \cite{Saleten-1971} in terms of the ordinary partial derivative of $L$ with respect to the inertial time derivative $\partial_0\xi_\mu$. Then, from expression (\ref{lagra}) and taking into account (\ref{sumand1}), (\ref{sumand2}) and (\ref{sumand3}), we obtain this general expression
\begin{equation}\label{moment2}
\Pi^\mu = 4\, \Big( a S^{0\mu} - b F^{0\mu} + 2\  c \ (\nabla \cdot \xi) g^{0\mu}\Big).
\end{equation}
For a geodesic and vorticity free unit observer, whose metrically equivalent 1-form is written as $u=-{\rm d}t=(-1,0,0,0)$, the conjugate momentum of a vector field $\xi$ relative to $u$ could be defined as
\begin{equation}\label{momentrel}
\Pi_u (\xi) = -4\, \Big(
a\ i(u) S - b \ i(u) F + 2\ c \ (\nabla \cdot \xi)\  u \Big).
\end{equation}
Note that, if $\mu=\nu=0$ ($\rho=0$), then $L$ is independent of the components of $\xi$.  In such a case, 
a study of the time-independent character of a volume integral quantity derived from $\Pi^\mu$ for an observer 
could be examined as a generalization of the treatment given in \cite{Saleten-1971}.


\section{Eikonal-like decomposition of the field equation}   
\label{sec:5}

In Physics, the eikonal description provides the framework for field wave propagation. Rooted on the Hamilton-Jacobi equation, it  describes the electromagnetic wave propagation within the geometric optics approximation.  In this approximation, the electromagnetic potential vector is decomposed into the product of an amplitude and an exponential term carrying a phase (referred to as `eikonal') that varies rapidly. This formalism can be adapted to address equation (\ref{Eq-camp-2}) by assuming a suitable field factorisation, which we term eikonal-like factorisation. No approximations are made in this section, whose primary aim is to explore the implications of such a factorisation. However, in a weakly perturbed Minkowski spacetime gravitational field (with the metric expressed as $g_{\mu\nu} =  \eta_{\mu\nu} + h_{\mu\nu}$ in conventional notation), the influence of a hypothetical vector field on the propagation of tensorial perturbations (gravitational waves) can be analysed for a self-gravitating field coupled to the Einstein equation, or in the context of a tensor-vector gravity \cite{wil93} and its extensions.

Without lost of generality, at each space-time event, the field $\xi$ may be factorised as a vector field, $A$, and a scalar function $f$ of $\varphi$,  which depend on the event coordinates, $\{x^\mu\}$, that is:
\begin{equation}\label{factor-xi}
\xi(x^\mu) = A(x^\mu) f(\varphi(x^\mu))\, .
\end{equation}
In the eikonal ansatz, the vector $A$ represents an  amplitude vector and the function $\varphi$ is taken as a phase. Here we will consider this factorization, in general, without being necessarily an amplitude vector and a phase.

From this factorization of the field, $\xi = f A$, we get that $\nabla\xi= df\otimes A + f\ \nabla A$, which implies
\begin{equation}\label{div}
    \nabla \cdot \xi=i(A) df + f\ \nabla\cdot A = f' (k\cdot A)+ f\ \nabla\cdot A
\end{equation}
by contraction, with $df= f' k$ being $f' = df/d\varphi$,  $k \equiv {\rm d} \varphi$, and $k \cdot A \equiv g(k, A)$. Moreover,
$$
\nabla\nabla \xi=\nabla df \otimes A + 2\ df\otimes \nabla A +f\ \nabla\nabla A,
$$
where $\nabla df=f'' k\otimes k + f'\nabla k$ and $f'' = df'/d\varphi$. By contraction, the above equation gives
\begin{equation}\label{lap}
\Delta \xi =(\Delta f)  A + 2\ i(df) \nabla A +f\ \Delta A
\end{equation}
for $\Delta f=g^{\mu\nu}\nabla_{\mu}\partial_\nu f=f'' k^2 +f'\nabla\cdot k$. Then, we can rewrite (\ref{lap}) as
\begin{equation}\label{lap2}
\Delta \xi =f\Delta A+ f'\Big( 2\nabla_k A+ (\nabla\cdot k) A\Big)+ f'' k^2\, A,
\end{equation}
with $k^2 \equiv g(k, k)$. And, from expression (\ref{div}),
\begin{equation}\label{div2}
    d(\nabla \cdot \xi)=f'' (k\cdot A) k + f'\Big(d(k\cdot A)+ (\nabla\cdot A) k\Big) + f\ d (\nabla\cdot A).
\end{equation}
Now, replacing these expressions (\ref{lap2}) and (\ref{div2}) in Eq. (\ref{Eq-camp-2}), we have the following results.

\begin{proposition}
Given the field $\xi$ factorised as $\xi=f A$ with $f$ a scalar function of $\varphi$ and $A$ a vector field, the field equation (\ref{Eq-camp-2}) is written as
\begin{equation}\label{descompo}
f {\cal X} + f' {\cal Y} + f'' {\cal Z} = 0,
\end{equation}
with $ {\cal X, Y, Z}$ defined by the expressions:
\begin{eqnarray}
{\cal X} &\equiv& \alpha \Delta A + \beta i(A) Ric + \gamma d(\nabla \cdot A) + \rho A \label{X}, \\ 
{\cal Y} &\equiv & \alpha[ 2 \nabla_k A + (\nabla \cdot k) A] + \gamma [{\rm d} (k \cdot A) + (\nabla \cdot A)\, k]\label{Y},  \quad  \,  \\ 
{\cal Z} &\equiv& \alpha k^2 \, A + \gamma (k \cdot A) \, k \label{Z},
\end{eqnarray}
where $k=d\varphi$ and $f'$ ($f''$) is the first (second) derivative of $f$ with respect to $\varphi$.
\end{proposition}

Moreover, the following statement has also been established. 
\begin{proposition}
For any given space-time metric $g$, a sufficient condition for a field $\xi = f A$ to satisfy Eq. (\ref{Eq-camp-2}) is that the vector $A$ and the differential of $\varphi$, $k=d\varphi$, obey the equations: 
\begin{equation}\label{eqsufi}
{\cal X}=0, \qquad {\cal Y}=0 , \qquad {\cal Z}=0 \, .
\end{equation}
with ${\cal X}, {\cal Y}$ and  ${\cal Z}$ defined by (\ref{X}), (\ref{Y}) and (\ref{Z}), respectively. 
\end{proposition}

Now, we analyse some interesting implications derived from Eqs. (\ref{eqsufi}). From the condition ${\cal Y} = 0$ we have that contraction $i(A){\cal Y} = 0$ holds, that is,
\begin{equation}
\begin{array}{ccl}
i(A){\cal Y} & = &  
\alpha[ 2 i(A)\nabla_k A + (\nabla \cdot k) A^2] \\ [0.2cm]
& + & \gamma [i(A){\rm d} (k \cdot A) + (\nabla \cdot A)\, (k\cdot A)]\\ [0.2cm]
& = & \nabla\cdot (\alpha A^2 k + \gamma (k \cdot A) A)= 0. 
\end{array}
\end{equation}
So, as a consequence of ${\cal Y} = 0$ we can define the  divergence free vector, ${\cal J}$, given by
\begin{equation}
 {\cal J} \equiv \alpha A^2 k + \gamma (k \cdot A) A, 
 \label{chi}
\end{equation}
that is, $\nabla\cdot {\cal J}=0$.

On the other hand, the contractions of (\ref{Z}) with $A$ and $k$ give:  
\begin{eqnarray}
i(A) {\cal Z} & = & \alpha k^2 A^2 + \gamma  (k \cdot A)^2 \label{AZ}, \\
i(k) {\cal Z} & = & (\alpha + \gamma) (k \cdot A) k^2 \label{kZ},
\end{eqnarray}
which are equal to zero when considering ${\cal Z}=0$. 
Analyzing these last equations ($i(A) {\cal Z}=0$ and $i(k) {\cal Z}=0$) we can distinguish the following cases:
\vspace{0.2cm}

\noindent \underline{Case 1:} $\alpha \gamma \neq 0$, $\alpha + \gamma \neq 0$. 
\vspace{0.2cm}

In this case, from expressions (\ref{AZ}) and (\ref{kZ}) equal to 0 we can conclude that $k \cdot A = 0$ and consequently $k$ and $A$ are orthogonal.
Note that if $k \cdot A \neq 0$, then $i(k) {\cal Z}=0$  
gives $k^2 = 0$ and $i(A) {\cal Z} = \gamma (k \cdot A)^2 = 0$, which enters in contradiction with $k \cdot A = 0$.

Consequently, from $k\cdot A=0$ we get the following results:
\begin{itemize}
\item[(i)] From Eq. (\ref{AZ}) we have $i(A){\cal Z} = \alpha k^2 A^2=0$ and then $k$ and $A$ are both null and collinear, or one of them is null and the other one is space-like. 

\item[(ii)] From Eq. (\ref{chi}), ${\cal J} = \alpha A^2 k$ is divergence free. Then, since $k$ is a gradient ($k = d \varphi$), the Hessian of $\varphi$, $\nabla d\varphi=\nabla k$, is symmetric and if $k$ is null,
we have that
$$
\nabla_k k=i(k)\nabla k=i(k) ^t\nabla k=\frac{1}{2}\nabla (k^2)=0
$$
and the integral curves of $k$ define an affine parameterised congruence of null geodesics. If, in addition, the amplitude  vector $A$ is space-like, then ${\cal J}$ is a conserved null four-current, $\nabla\cdot(A^2 k) = 0$. 

\item[(iii)] By using (\ref{Y}), the equation ${\cal Y} = 0$ simplifies to
\begin{equation} \label{Ysim}
\alpha[ 2 \nabla_k A + (\nabla \cdot k) A] + \gamma (\nabla \cdot A)\, k = 0\, .
\end{equation}
\end{itemize}

\noindent \underline{Case 2:} $\alpha \gamma \neq 0$, $\alpha + \gamma = 0$. 
\vspace{0.2cm}

In this case,  from (\ref{AZ}) equal to 0 we have that:
\begin{equation}\label{AZ2}
i(A) {\cal Z} = k^2 A^2 - (k \cdot A)^2 = 0.
\end{equation}
Then, we get the following options: $A$ and $k$ are both null and collinear; or they generate a null 2-plane, that is, one of them is null and the other one is space-like or both vectors are space-like and non orthogonal, that is, $k^2 A^2 = (k \cdot A)^2 \neq 0$.

Notice that, in both analysed cases, neither $k$ nor $A$ can be time-like. As a consequence, we have the following statement:

\begin{proposition}
In the generic case, i.e. when $\alpha \gamma \neq 0$, the eikonal-like decomposition, $\xi=f \, A$, satisfying (\ref{eqsufi}), 
is not possible for a time-like vector field $\xi$. 
\end{proposition}

\noindent \underline{Case 3:} $\alpha \gamma = 0$. 
\vspace{0.2cm}

In this case, if $\alpha=0$ and $\gamma\neq 0$, from ${\cal Z}=0$ we get that $k$ and $A$ are orthogonal ($k\cdot A=0$). Moreover, the equation ${\cal Y}=0$ reduces to $\nabla \cdot A=0$ and ${\cal X}=0$ implies that $A$ is an eigenvector of the Ricci tensor.

If $\gamma=0$ and $\alpha\neq 0$ then, from ${\cal Z}=0$, the vector $k$ is null. Moreover, the contraction $i(A){\cal Y}=0$ implies that 
${\cal J}=\alpha A^2 k$, with $\nabla\cdot {\cal J}=0$. 

\vspace{0.3cm}

The implications derived from a sufficient condition for an eikonal-like factorization of the field $\xi$, $\xi=f\, A$, have been exhaustively examined. Notice that the Proca field equation (\ref{Proca-2}) belongs to case 2 since $\alpha=1$ and $\gamma=-1$. However, the AKE equation (\ref{AKE}), the Will one (\ref{Will-2-BoHa}) and the Dale-S\'aez equation (\ref{DS-eq}) are in any of the cases depending on the value of the coefficients.

%

\section{Vector field energy-momentum tensor.}   
\label{sec:4}


\subsection{The Hilbert energy tensor: generalities.}
\label{sec:4a}

The Hilbert energy tensor, $T$, is obtained from the variation of the  functional (\ref{funcional}) with respect to the metric $g$    (see, for instance, Refs. \cite{Hawking-Ellis,gu2021,mis17}). As starting point to obtain $T$, let us consider the contravariant expression:
\begin{equation}\label{T-Hilbert}
T^{\mu\nu} = \frac{2}{\sqrt{- {\rm g}}} \frac{\delta {\cal L}}{\delta g_{\mu\nu}}, \qquad {\cal L} \equiv L \sqrt{- {\rm g}}, 
\end{equation}
with $L$ the considered quadratic Lagrangian density (\ref{lagra}). Here, the function ${\cal L}$ depends at most on the metric field and its first derivatives, ${\cal L} (g_{\mu\nu}, g_{\mu\nu,\rho})$, so that the variational derivative takes the expression:
\begin{equation}\label{delta-calL}
\frac{\delta {\cal L}}{\delta g_{\mu\nu}} = \frac{\partial {\cal L}}{\partial g_{\mu\nu}} - 
\frac{\partial}{\partial x^\rho} \frac{\partial {\cal L}}{\partial g_{\mu\nu,\rho}}
\end{equation}
when the variations $\delta g_{\mu\nu}$ and $\delta g_{\mu\nu,\rho}$ vanish on the integration boundary. 

Since no metric derivative appears in $\rm{g}$, the development of the second member of (\ref{delta-calL}) leads to
\begin{eqnarray}
\frac{\delta{\cal L}}{\delta{g_{\mu\nu}}} & = & \sqrt{-\rm{g}} \Big( \frac{\partial L}{\partial{g_{\mu\nu}}} - \partial_\rho  \frac{\partial L}{\partial g_{\mu\nu,\rho}}\Big)+ L \, \frac{\partial\sqrt{-\rm{g}}}{\partial{g_{\mu\nu}}} - (\partial_\rho \sqrt{-\rm{g}}) \frac{\partial L}{\partial g_{\mu\nu,\rho}} \nonumber \\ 
& = & \sqrt{-\rm{g}} \frac{\delta L}{\delta{g_{\mu\nu}}} + L \, \frac {1}{2 \sqrt{-\rm{g}}} \frac{\partial (- \rm{g})}{\partial g_{\mu\nu}} \nonumber
- (\partial_\rho \sqrt{-\rm{g}}) \frac{\partial L}{\partial{g_{\mu\nu,\rho}}},  
\end{eqnarray}
and thus, taking into account (\ref{lng}), this relevant expression for the Hilbert energy tensor follows (cf. \cite{gu2021})
\begin{equation}\label{T-Hilbert-bis}
T^{\mu\nu} = 2 \frac{\delta L}{\delta g_{\mu\nu}} + L g^{\mu\nu} - 2 \Gamma^\alpha_{\alpha\rho} \frac{\partial L}{\partial g_{\mu\nu,\rho}}
\end{equation}
where the first term is the variational derivative of  the Lagrangian density $L$,  that is%
\begin{equation}\label{delta-L}
\frac{\delta {L}}{\delta g_{\mu\nu}} = \frac{\partial {L}}{\partial g_{\mu\nu}} - 
\frac{\partial}{\partial x^\rho}  \frac{\partial {L}}{\partial g_{\mu\nu,\rho}} \, .
\end{equation}
Different expressions of $L$ will give its corresponding energy tensors $T$. 
In our case, we will consider $L$ given by (\ref{lagra}) and analyze it term by term separately to obtain the corresponding energy tensor $T$ by addition of each one of the studied parts.


\subsection{Hilbert energy tensor associated to the Lagrangian density of $\xi$.}
\label{sec:4b}

In this subsection, we will focus on the Lagrangian density (\ref{lagra}) in order to derive the Hilbert energy tensor using Eq. (\ref{T-Hilbert-bis}). As we have mentioned before, we will analyze term by term the factors appearing in $L$.

To begin with, let us consider a Lagrangian density defined by a twice differentiable function of $\nabla \cdot \xi$, say $L \equiv h(\nabla \cdot \xi)$. To obtain the Hilbert Energy tensor corresponding to this $L$, we take into account that the divergence of a vector field $\xi$ can be conveniently expressed as:
\begin{eqnarray}\label{divxi}
\nabla \cdot \xi & = & g^{\alpha \beta} \nabla_\alpha \xi_\beta = g^{\alpha \beta} (\partial_\alpha \xi_\beta - \Gamma^\lambda_{{\alpha \beta}} \xi_\lambda) \nonumber \\  & = & g^{\alpha \beta} \partial_\alpha \xi_\beta - g^{\alpha \beta} g^{\lambda \sigma} \Gamma_{\alpha \beta .  \sigma} \xi_\lambda \, , 
\end{eqnarray}
and, by derivation, it leads to
\begin{eqnarray}
2 \, \frac{\partial (\nabla \cdot \xi)} {\partial g_{\mu\nu}} & = & g^{\alpha \beta}
(\Gamma^\mu_{{\alpha \beta}} \xi^\nu + \Gamma^\nu_{{\alpha \beta}} \xi^\mu ) - ({\cal L}_\xi g)^{\mu\nu} \, ,\label{par-div} \\  
2 \, \frac{\partial (\nabla \cdot \xi)} {\partial g_{\mu\nu,\rho}} & = & g^{\mu\nu} \xi^\rho -  
g^{\rho\mu} \xi^\nu - g^{\rho\nu}\xi^\mu\, . \label{par-div-der}
\end{eqnarray}
Then, using expression (\ref{divxi}), its partial derivative and the definition of variational derivative given in (\ref{delta-L}), we have the following result:
\begin{lemma}
The variational derivative of $\nabla \cdot \xi$ with respect to metric variations is given by:
\begin{eqnarray}\label{delta-div}
\frac{\delta (\nabla \cdot \xi)}{\delta g_{\mu\nu}} & = & \frac{\partial (\nabla \cdot \xi)}{\partial g_{\mu\nu}} - 
\frac{\partial}{\partial x^\rho}  \frac{\partial (\nabla \cdot \xi)}{\partial g_{\mu\nu,\rho}} \nonumber \\ & = & 
- \frac{1}{2} \big [g^{\mu\nu} \partial_\rho \xi^\rho + (g^{\mu\beta} \xi^\nu + g^{\nu\beta} \xi^\mu)\Gamma^\rho_{\beta\rho} \big]. \quad
\end{eqnarray}
\end{lemma}

Now, let us obtain the Hilbert energy tensor associated with $L \equiv h(\nabla \cdot \xi)$. Eq. (\ref{T-Hilbert-bis}) applied to this case gives:
\begin{equation}\label{T-Hilbert-div1}
T^{\mu\nu} = 2 h' \frac{\delta (\nabla \cdot \xi)}{\delta g_{\mu\nu}}  + h g^{\mu\nu} 
- 2 \big[ h'' \partial_\rho (\nabla \cdot \xi) + h' \Gamma^\alpha_{\alpha\rho}\big] \frac{\partial L}{\partial g_{\mu\nu,\rho}}
\end{equation}
where the prime denotes derivative with respect to $\nabla \cdot \xi$. 

Finally, by direct substitution of (\ref{par-div}) and (\ref{par-div-der}) in Eq. (\ref{T-Hilbert-div1}),  we get:
\begin{lemma}
Given a vector field $\xi$, the Hilbert tensor associated with the Lagrangian density $L = h(\nabla \cdot \xi)$ is%
\begin{equation}\label{T-Hilbert-div2}
T = \big[ h - (\nabla \cdot \xi) h' - h'' i(\xi) {\rm d} (\nabla \cdot \xi)\big] \, g + h'' \xi \tilde{\otimes} d (\nabla \cdot \xi),
\end{equation}
where $h'$ and $h''$ stand for the first and second derivative of $h$ with respect $\nabla \cdot \xi$, respectively.
\end{lemma}
\vspace{0.2cm}

By applying this results to the interesting particular cases $L =\nabla \cdot \xi$ and $L=( \nabla \cdot \xi)^2$, we can conclude the following statements.
\begin{proposition}\label{L1}
 If $L = \nabla \cdot \xi$, then $T = 0$.\\
\end{proposition}

\begin{proposition}
If $L = (\nabla \cdot \xi)^2$, then
\begin{equation}  \label{Tdiv2}
T = - [(\nabla \cdot \xi)^2 + 2 \,  i(\xi){\rm d} (\nabla \cdot \xi)] \, g + 2 \, \xi \tilde{\otimes} {\rm d} (\nabla \cdot \xi).
\end{equation}   
\end{proposition}

Notice that Proposition \ref{L1} agrees with the fact that terms proportional to $2 \nabla\cdot\xi = \tr S$ do not contribute to neither the energy tensor nor the field equations.

Next, let us consider the case $L = \tr F^2$ with $F=d\xi$. From (\ref{trF2}), $\tr F^2=-F_{\mu\nu}F^{\mu\nu}=-g^{\mu\alpha}g^{\nu\beta}F_{\mu\nu}F_{\alpha\beta}$ with
\begin{equation}\label{F-comp}
F_{\alpha\beta} = \partial_\alpha \xi_\beta - \partial_\beta \xi_\alpha .
\end{equation}
Since no metric derivative occurs in $L$, we have:
\begin{equation}\label{partialF2}
\frac{\partial \tr F^2}{\partial g_{\mu\nu,\rho}} = 0, 
\end{equation}
and, from (\ref{delta-L}), its variational derivative with respect to the metric reduces to
\begin{equation}\label{delta-trF2}
\frac{\delta \tr F^2}{\delta g_{\mu\nu}} =  \frac{\partial \tr F^2}{\partial g_{\mu\nu}} = - 2 F^\mu_{\, \, \, \alpha} F^{\alpha \nu} .
\end{equation}
In this case, the calculation of the Hilbert energy tensor from Eq. (\ref{T-Hilbert-bis}) gives
\begin{equation}\label{T-Hilbert-trF2}
T^{\mu\nu} = - 4(F^2)^{\mu \nu} + (\tr F^2) \, g^{\mu\nu},  
\end{equation}
and we can conclude the following statement:
\begin{proposition}
 If $L =  \frac{1}{4} \tr F^2$,  then 
\begin{equation}\label{TtrF2}
T = - F^2 + \frac{1}{4} (\tr F^2) \, g,
\end{equation}
where the 1/4 factor has been included for convenience.
\end{proposition}

Note that, in this last statement, $F$ can represent the electromagnetic field in a curved space-time, recovering the known result in flat space-time and extending it to an arbitrary space-time without considering the equivalence principle.

Incidentally, the contribution to $T$ of a $\xi^2$ term in the Lagrangian density $L$ is given by $-2\xi\otimes\xi +\xi^2 g$ (where the minus sign comes from the variation of $\xi^2=g^{\alpha\beta}\xi_{\alpha}\xi_{\beta}$ with respect to $g_{\mu\nu}$). Then, this known result is recovered:
\begin{proposition}
For $L_{\cal P} = \displaystyle\frac{1}{4}\tr F^2 - \frac{\mu_p^2}{2} \, \xi^2$, being the Proca Lagrangian density, the associated Hilbert energy tensor $T_{\cal P} $ is given by
\begin{equation}\label{Tproca}
T_{\cal P} =-F^2+\frac{1}{4}(\tr F^2)\ g+\mu_p^2\left(\xi\otimes\xi-\frac{1}{2}\xi^2 g\right)
\end{equation}
where $\mu_p$ is the Proca mass.
\end{proposition}

Now, we consider the Lagrangian density $L=\tr S^2$ with $S={\cal L}_\xi \, g$. In components, this Lagrangian density has the expression (\ref{trS2}) with
\begin{equation} \label{S-comp}
S_{\alpha\beta} =  \partial_\alpha \xi_\beta + \partial_\beta \xi_\alpha - g^{\rho\sigma} (g_{\alpha\sigma,\beta} + g_{\beta\sigma,\alpha} - g_{\alpha\beta,\sigma}) \xi_\rho .
\end{equation}
In this case calculations are a bit more involved than in the previous case since, from  Eqs. (\ref{T-Hilbert-bis}) and (\ref{delta-L}), the Hilbert tensor is written as
\begin{equation}\label{T-Hilbert-trS^2}
T^{\mu\nu} =   (\tr S^2) g^{\mu\nu} + 2 \,  \frac{\partial \tr S^2}{\partial g_{\mu\nu}} - 
2 \, \partial_\rho \frac{\partial \tr S^2}{\partial g_{\mu\nu,\rho}} 
-  2 \, \Gamma^\alpha_{\alpha\rho} \frac{\partial \tr S^2}{\partial g_{\mu\nu,\rho}} \, ,\, 
\end{equation}
and the following expressions 
 \begin{eqnarray}
\frac{\partial \tr S^2} {\partial g_{\mu\nu}} & = & - 2 (S^2)^{\mu\nu} + 
2 S^{\alpha\beta}(\xi^\mu \Gamma^\nu_{\alpha\beta} + \xi^\nu \Gamma^\mu_{\alpha\beta}),\label{par-trS2} \quad \\  
\frac{\partial \tr S^2} {\partial g_{\mu\nu,\rho}} & = & - 2 (\xi^\mu S^{\nu\rho} + \xi^\nu S^{\mu\rho} - \xi^\rho S^{\mu\nu}), \label{par-trLS2-der}
\end{eqnarray}
have to be taken into account. After substitution of (\ref{par-trS2}) and (\ref{par-trLS2-der}) in (\ref{T-Hilbert-trS^2}), and after some rearrangements of terms and simplifications, we obtain the following expression for the energy tensor:
\begin{eqnarray}\label{T-Hilbert-trS^2-bis}
T^{\mu\nu} & = &  \tr S^2 g^{\mu\nu} -  4 (S^ 2)^{\mu\nu} - 4 (\partial_\rho \xi^\rho + \Gamma^{\alpha}_{\alpha\rho}) \,  \xi^\rho S^{\mu\nu} \nonumber \\
&  &+ 4 \, \xi^\mu (\partial_\rho S^{\rho \nu} + \Gamma^\alpha_{\alpha\rho}S^{\rho \nu} + \Gamma^\nu_{\alpha\beta}S^{\alpha \beta})\nonumber\\
& & + 4 \, \xi^\nu (\partial_\rho S^{\rho \mu} + \Gamma^\alpha_{\alpha\rho}S^{\rho \mu} + \Gamma^\mu_{\alpha\beta}S^{\alpha \beta})\nonumber\\
& & -  4 \, (\xi^\rho \partial_\rho S^{\mu\nu} - S^{\mu\rho} \partial_\rho \xi^\nu - S^{\rho \nu} \partial_\rho \xi^\mu)\, .
\end{eqnarray}
This expression involves the divergence of $\xi$ at the third term, the divergence of $S$ at the fourth and fifth terms and the Lie derivative of the contravariant tensor $S^{\mu\nu}$ along $\xi$  at the last term. Then, we can conclude the following statement:
\begin{proposition}
If $L = \frac{1}{4} \tr S^2$,  then
\begin{equation}\label{TtrS2}
T = - S^2 + \frac{1}{4} (\tr S^2) \, g - (\nabla \cdot \xi) S + 
\xi \tilde{\otimes} (\nabla \cdot S) - {\cal L}_\xi S^*,
\end{equation}
where $S^*$ represents the contravariant tensor metrically equivalent to $S = {\cal L}_\xi \, g$. That is, $S^* = ({\cal L}_\xi g)^* = - {\cal L}_\xi g^*$, where $g^*$ is the contravariant metric tensor.     
\end{proposition}
 
As usual,  where expressions in components are used, the star in $S^*$ and $g^*$ will be tacitly understood, that is, $(S^*)^{\mu\nu} = g^{\mu\alpha} g^{\nu \beta} S_{\alpha \beta} \equiv S^{\mu\nu}$. Thus, the expression for the last term in (\ref {TtrS2}), ${\cal L}_\xi S^*$, is written as 
\begin{eqnarray}\label{Lie2Scontra}
({\cal L}_\xi S^*)^{\mu\nu} & = & \xi^\rho \partial_\rho S^{\mu\nu} - S^{\mu\rho} \partial_\rho \xi^\nu - S^{\rho \nu} \partial_\rho \xi^\mu \nonumber \\ 
& = & \xi^\rho \nabla_\rho S^{\mu\nu} - S^{\mu\rho} \nabla_\rho \xi^\nu - S^{\rho \nu} \nabla_\rho \xi^\mu
\end{eqnarray}
where substitution of partial derivation by the covariant one becomes from the symmetry property of the Christoffel symbols. Again, the 1/4 factor in $L$ was included for convenience.

Incidentally, notice that the sum of the first and the last terms in Eq. (\ref{TtrS2}),  might be conveniently replaced from the following identity:
\begin{equation}\label{S2Lie}
S^2 + {\cal L}_\xi S^* = \nabla_\xi S +  \nabla \xi \times \vspace{0.5mm}  ^{\rm t}\nabla \xi - \vspace{0.5mm} ^{\rm t}\nabla \xi \times \nabla \xi \, , 
\end{equation}
where each summand is a symmetric tensor. 
Notice that Eqs. (\ref{trsi}) and (\ref{trsi2}) can be used to deduce Eq. (\ref{S2Lie}).
Once this relation between symmetric contravariant tensors has been established, the expression remains valid for the corresponding  covariant, or mixed,  metrically equivalent tensors, due to its tensorial character. A similar situation occurs in Eqs. (\ref{Tdiv2}), (\ref{TtrF2}) and (\ref{Tproca}) of the Hilbert tensor in the corresponding statements. 

Finally, we comment on the whole Hilbert energy-momentum tensor corresponding to the Lagrangian (\ref{lagra}), introducing a constant $1/4$ factor for notational convenience. This factor does not alter the expression of the field equations (\ref{Eq-camp-1}) nor (\ref{Eq-camp-2}), but cancels the presence of the factor $4$ in the expression for $T$. The Hilbert energy tensor derived from the whole Lagrangian density $\frac{1}{4} \sqrt{- {\rm g}} \, L$, with $L$  expressed as (\ref{lagra}) with $\nu=0$, is given by:
\begin{eqnarray} 
T & = & a \,T_{\tiny I} + b \,T_{\tiny II} + c \,T_{\tiny III} + \mu \, T_{\tiny IV} \nonumber
\end{eqnarray}
where
\begin{eqnarray} \nonumber
T_{\tiny I} & = & - \nabla_\xi S - \nabla \xi \times ^t \hspace{-1mm} \nabla \xi + ^t \hspace{-1mm}\nabla \xi \times \nabla \xi - (\nabla \cdot \xi) S + 
\xi {\tilde\otimes} \nabla \cdot S + \frac{1}{4} ({\rm tr} S^2) g \nonumber \\
T_{\tiny II} & = & - F^2 +  \frac{1}{4} ({\rm tr} F^2) g\nonumber \\
T_{\tiny III} & = & 2 \xi {\tilde\otimes} d (\nabla \cdot \xi) - \big[ (\nabla \cdot \xi)^2 + \frac{1}{2} i(\xi) d (\nabla \cdot \xi) \big] g \nonumber \\
T_{\tiny IV} & = & - \frac{1}{2} \xi \otimes \xi +  \frac{1}{4} \xi^2 g \nonumber 
\end{eqnarray}
The above expressions summarise the results obtained, step by step, in this section. The subindexes $I$,  $II$, $III$ and $IV$ denote the energy tensor contributions from the first $(\frac{1}{4}{\rm tr} S^2)$, second $(\frac{1}{4}{\rm tr} F^2)$, third $(\frac{1}{4}({\rm tr} S)^2 = (\nabla \cdot \xi)^2)$ and fourth $(\frac{1}{4} \xi^2)$  Lagrangian pieces, respectively. 

Note that the different $T$-summands which are proportional to the metric are irrelevant for the algebraic classification of the total energy tensor $T$. Nevertheless, these terms may  be significant in a cosmological scenario involving a self-gravitating field coupled  to Einstein gravity or in the context of non-Abelian or extended vector-tensor gravity theories, as mentioned in the Introduction.

%
%

\section{Summary and discussion}   
\label{sec:6}

In this work we have examined the dynamics of a relativistic vector field $\xi$ in a curved space-time. We have derived, from variational principles, the field equations and the energy content in any metric $g$ coupled to the field from a quadratic Lagrangian field density and keeping at most the first order field derivatives. The field equation (\ref{Eq-camp-2}) reduces to the one considered by other authors when the general expression is particularised using a properly parameter trial. The associated energy tensor has been obtained analyzing the Lagrangian terms separately and in a general way, which can be useful to characterize it algebraically. Its whole expression has been given at the end of Section 5. Moreover, this presentation will also allow to perform a $3+1$ decomposition of both, the field equation and the energy tensor, to interpret them from an observer point of view.

Furthermore, the eikonal-like ansatz has also been studied, concluding that, a separability hypothesis does not apply when $\alpha \gamma \neq 0$ and $\xi$ is time-like. 
 
It is worthwhile to remark that the addition of a term proportional to $\tr F^3$ is irrelevant since
$$
\frac{\partial \tr F^3}{\partial \nabla_\alpha \xi_\beta} = 0.
$$
However, a term proportional to $\tr S^3$ leads to second order derivatives of the field in the field equations since
$$
\frac{\partial \tr S^3}{\partial \nabla_\alpha \xi_\beta} = 6 (S^2)^{\alpha\beta} ,
$$
and $(\nabla\cdot S^2)_\mu=
(\nabla_\alpha S^{\alpha\beta})S_{\beta\mu} + S^{\alpha\beta}(\nabla_\alpha S_{\beta\mu}),
$
which can be developed taking into account the definition (\ref{def-S}) of $S$. 

This study has been developed using the Levi-Civita connexion. However, it admits an extension by considering a general linear connexion (with torsion and non-metricity tensor contributions) and following the approach presented in \cite{Lavinia2019} but using the methodology here presented. Nevertheless, this task is out of the scope of the present work.

On the other hand, in Electrodynamics, the Lorenz gauge condition on the vector potential $\xi$, $\nabla \cdot \xi = 0$, not only simplifies the study of Maxwell equations in Minkowski space-time, but also,  in a curved  background with metric $g$. Nevertheless, in the last situation, the vector potential $\xi$ couples to the Ricci tensor of the metric $g$, $Ric(g)$, via an algebraic term of the form $i(\xi) Ric(g)$. This term could have a relevant contribution for non vacuum geometries. By contrast, if $\xi$ represents the Proca vector potential, the equation $\nabla \cdot \xi = 0$ becomes from the Proca field equation, and plays the role of a constraint equation (not of a gauge condition). Now, a lineal  term proportional to the own potential $\xi$ has to be considered, in addition to the four dimensional Laplace-Beltrami curved contribution, say $\Delta \xi$,  and the gravitational coupling term $i(\xi) Ric(g)$.

The natural generalization from Electrodynamics is to assume that $L$ is a scalar density made from quadratic terms in the field and its first derivatives of first order at most. In this case, the vanishing of the divergence of $\xi$ cannot be given for granted, so that its differential, $d(\nabla \cdot \xi)$, enters in the $\xi$ field equations. The general stress-energy tensor $T$ of the $\xi$ field, has been obtained by taking the variational derivative of $L$ with respect to the space-time metric $g$. This is a topic that should be complemented with the algebraic classification of $T$ and its physical interpretation.

The above considerations could be taken into account to go further with the study here presented. At least, three scenarios could be studied:  (i) $\xi$ is a dynamic test field in a particular (in general, curved) known space-time geometry;
(ii) $\xi$ is a self-gravitating field dynamically coupled to the Einstein field equations; and (iii) $\xi$ is, jointly the space-time geometry, a dynamical vector field whose field equation couples to the extended Einstein equations for a vector-tensor gravity theory, and whose energy tensor has to be added at the Einstein tensor and provided with a (purely) geometric meaning. 

Nonetheless, whether two solutions of the field equations for  the same $\xi$ in (ii) and (iii) are equivalent (that is,  a simple matter of interpretation) or whether they are necessarily not physically equivalent, is an issue that requires to be clarified in depth.  Meanwhile, applications of vector tensor theories of the gravitation (such as the search for exact solutions or the description of gravitational scenarios in presence of a vector field) is an interesting field to gain progress on this issue.



\bmhead{Acknowledgements}

This work has been supported by the Spanish Ministerio
de Ciencia, Innovaci\'on y Universidades, Projects:
PID2019-109753GB-C21/AEI/10.13039/501100011033 and
PID2019-109753GB-C22/AEI/10.13039/501100011033; and by the European Union Next Generation EU/PRTR, Programa de Planes Complementarios I+D+I, ref. ASFAE/2022/014.
All three authors were also financially supported by Generalitat Valenciana, Spain, grant CIAICO/2022/252.

\section*{Declarations}

\begin{itemize}
\item Funding

This work has been supported by the Spanish Ministerio
de Ciencia, Innovaci\'on y Universidades, Projects:
PID2019-109753GB-C21/AEI/10.13039/501100011033 and
PID2019-109753GB-C22/AEI/10.13039/501100011033; and by the European Union Next Generation EU/PRTR, Programa de Planes Complementarios I+D+I, ref. ASFAE/2022/014.
All three authors were also financially supported by Generalitat Valenciana, Spain, grant CIAICO/2022/252.

\item Conflict of interest

The authors certify that there are no conflicts of interest.

\item Data availability

Not applicable
 
\item Materials availability

Not applicable

\item Code availability 

Not applicable
\item Author contribution

All the authors contributed equally to this work.
\end{itemize}

\noindent
If any of the sections are not relevant to your manuscript, please include the heading and write `Not applicable' for that section.



\end{document}